\documentstyle[preprint,aps]{revtex}
\begin{document}
\title{Experimental reversion of the optimal quantum cloning and flipping processes}
\author{Fabio Sciarrino, Veronica Secondi, and Francesco De Martini}
\maketitle

\begin{abstract}
The quantum cloner machine maps an unknown arbitrary input qubit into two
optimal clones and one optimal flipped qubit. By combining linear and
non-linear optical methods we experimentally implement a scheme that, after
the cloning transformation, restores the original input qubit in one of the
output channels, by using local measurements, classical communication and
feedforward. This significant teleportation-like method demonstrates how the
information is preserved during the cloning process. The realization of the
reversion process is expected to find useful applications in the field of
modern multi-partite quantum cryptography.
\end{abstract}

\pacs{}

The conservation of information can be assumed as a basic principle of
physics \cite{Horo04}. Accordingly, since two perfect copies of an unknown
arbitrary quantum state $\left| \phi \right\rangle $ carry more information
than the latter one, it is impossible to realize a perfect quantum cloning
machine \cite{Woot82}.\ On the same token, since more information about $%
\left| \phi \right\rangle $ can be extracted from a pair of two orthogonal
qubits $\left| \phi \right\rangle \left| \phi ^{\perp }\right\rangle $ than
from two parallel ones $\left| \phi \right\rangle \left| \phi \right\rangle $
the impossibility argument holds for the spin flipping operation, i.e. the
NOT gate \cite{Bech99}. Even if the cloning and flipping operations are
unrealizable in their exact forms, they can be approximated ''optimally'',
i.e. with the minimum added noise, by the corresponding universal quantum
machines, i.e., the universal optimal quantum cloning machine (UOQCM) \cite
{Buze96} and the universal-NOT (U-NOT) gate \cite{Bech99,Gisi99}. In the
present paper we experimentally address the important problem whether, in
spite of their fundamental limitation, these optimal machines do conserve
all the information associated with any input qubit $\left| \phi
\right\rangle $ and how this information can be fully retrieved by a
teleportation-like scheme. More precisely, here the quantum information
content of $\left| \phi \right\rangle $ is spread over the 3 qubits
entangled state $\left| \Sigma (\phi )\right\rangle _{SAB}$ through the
quantum cloner and then is fully retrieved on an output channel by
implementing a Local Operation and Classical Communications (LOCC) method 
\cite{Brus01}. The spreading of the initial information over $\left| \Sigma
(\phi )\right\rangle _{SAB}$ is obtained by a simultaneous implementation of
the $1\rightarrow 2$ UOQCM and the $1\rightarrow 1$ U-NOT gate via a quantum
injected optical parametric amplifier $(QI-OPA)$ \cite{DeMa02,Lama02} or an
all linear optics setup \cite{Ricc04}. As shown in Figure 1, at the input of
UOQCM the three modes $S$, $A$, $B$ support respectively the qubit $\left|
\phi \right\rangle $ and the two ancillas. The quantum reversion process is
completed by application to the output of UOQCM\ of a Local Operation and
Classical Communication ($LOCC$) procedure \cite{Greg03}, indeed a modified
teleportation protocol consisting of a Bell measurement on the modes $S$ and 
$A$, a classical communication channel, and of a final unitary operation on
the qubit $B$ \cite{Benn96}.

Let us consider the cloning machine ($CM$) which realizes simultaneously the 
$1\rightarrow 2$ UOQCM and the $1\rightarrow 1$ U-NOT\ gate. We start from
the input qubit $\left| \phi \right\rangle \equiv \left| \phi \right\rangle
_{S}=\alpha \left| 0\right\rangle _{S}+\beta \left| 1\right\rangle _{S}$ and
the ancilla qubits $A$ and $B$ in the state $\left| 0\right\rangle $. After
the cloning process the overall output state $\left| \Sigma (\phi
)\right\rangle _{SAB}$ reads 
\begin{equation}
\left| \Sigma (\phi )\right\rangle _{SAB}=\sqrt{\frac{2}{3}}\left| \phi
\right\rangle _{S}\left| \phi \right\rangle _{A}\left| \phi ^{\perp
}\right\rangle _{B}-\frac{1}{\sqrt{6}}\left( \left| \phi \right\rangle
_{S}\left| \phi ^{\perp }\right\rangle _{A}+\left| \phi ^{\perp
}\right\rangle _{S}\left| \phi \right\rangle _{A}\right) \left| \phi
\right\rangle _{B}  \label{cloningout}
\end{equation}
The qubits $S$ and $A$ end up in the state $\rho _{S}=\rho _{A}=\frac{5}{6}%
\left| \phi \right\rangle \left\langle \phi \right| +\frac{1}{6}\left| \phi
^{\perp }\right\rangle \left\langle \phi ^{\perp }\right| $ and are the
optimal clones of the input qubit, while the qubit $B$, the optimally
flipped of $\left| \phi \right\rangle $, is found in the output state $\rho
_{B}=\frac{1}{3}\left| \phi \right\rangle \left\langle \phi \right| +\frac{2%
}{3}\left| \phi ^{\perp }\right\rangle \left\langle \phi ^{\perp }\right| $.
The state (\ref{cloningout}) can be interpreted as the quantum superposition
of two three-qubit entangled states, which depends on the complex parameters 
$\alpha $ and $\beta $: 
\begin{equation}
\left| \Sigma (\phi )\right\rangle _{SAB}=\alpha \left| \Sigma
(0)\right\rangle _{SAB}+\beta \left| \Sigma (1)\right\rangle _{SAB}
\label{Wstate}
\end{equation}
where $\left| \Sigma (0)\right\rangle =(2/3)^{-1/2}\left| 0\right\rangle
_{S}\left| 0\right\rangle _{A}\left| 1\right\rangle _{B}-6^{-1/2}\left(
\left| 0\right\rangle _{S}\left| 1\right\rangle _{A}\left| 0\right\rangle
_{B}+\left| 1\right\rangle _{S}\left| 0\right\rangle _{A}\left|
0\right\rangle _{B}\right) $ and $\left| \Sigma (1)\right\rangle
=(2/3)^{-1/2}\left| 1\right\rangle _{S}\left| 1\right\rangle _{A}\left|
0\right\rangle _{B}-6^{-1/2}\left( \left| 1\right\rangle _{S}\left|
0\right\rangle _{A}\left| 1\right\rangle _{B}+\left| 0\right\rangle
_{S}\left| 1\right\rangle _{A}\left| 1\right\rangle _{B}\right) $. These are
W three-qubit entangled states which are recognized to exhibit the highest
robustness of bipartite entanglement against the loss of one qubit \cite
{Dur00}, \cite{Brub02}, \cite{Eibl04}.

{\bf \ }The procedure adopted to reverse the cloning and flipping processes\
consists of a $LOCC$ approach similar to the quantum teleportation protocol,
as said \cite{Benn93}. To understand how the restoring of the initial qubit
is obtained, the state (\ref{cloningout}) can be re-expressed introducing
the Bell states of the qubits $S$ and $A:$ $\left| \Phi ^{\pm }\right\rangle
_{SA}=2^{-1/2}\left( \left| 0\right\rangle _{S}\left| 0\right\rangle _{A}\pm
\left| 1\right\rangle _{S}\left| 1\right\rangle _{A}\right) $ and $\left|
\Psi ^{\pm }\right\rangle _{SA}=2^{-1/2}\left( \left| 0\right\rangle
_{S}\left| 1\right\rangle _{A}\pm \left| 1\right\rangle _{S}\left|
0\right\rangle _{A}\right) .$ The cloner output state can hence be recast as 
\begin{equation}
\left| \Sigma (\phi )\right\rangle _{SAB}=\frac{1}{\sqrt{3}}\left[ \left|
\Phi ^{+}\right\rangle _{SA}i\sigma _{Y}\left| \phi \right\rangle
_{B}+\left| \Phi ^{-}\right\rangle _{SA}\sigma _{X}\left| \phi \right\rangle
_{B}+\left| \Psi ^{+}\right\rangle _{SA}\sigma _{Z}\left| \phi \right\rangle
_{B}\right]  \label{output function}
\end{equation}
We note that only the symmetric Bell states of $S$ and $A$ appear since the
two clones belong to the symmetric subspace.

Let us now describe the restoring machine ($RM$). For this purpose we
introduce two partners: Alice (${\cal A}$) and Bob (${\cal B}$) (Fig.1).
Alice holds the qubits $S$ and $A$ while Bob holds the qubit $B$. Alice
performs a Bell measurement on the qubits $S$ and $A,$ that is in the basis $%
\left\{ \left| \Phi ^{\pm }\right\rangle _{SA},\left| \Psi ^{\pm
}\right\rangle _{SA}\right\} $, and communicates the measurement result to
Bob sending a classical trit through a classical channel. Depending on
Alice's communication \cite{Giac02}, Bob applies a suitable Pauli operator $%
\sigma _{i}$ according to the following table: 
\begin{equation}
\begin{tabular}{||l||l||}
\hline\hline
Alice's result & Bob's operation \\ \hline\hline
$\left| \Phi ^{+}\right\rangle _{SA}$ & $i\sigma _{Y}$ \\ \hline\hline
$\left| \Phi ^{-}\right\rangle _{SA}$ & $\sigma _{X}$ \\ \hline\hline
$\left| \Psi ^{+}\right\rangle _{SA}$ & $\sigma _{Z}$ \\ \hline\hline
\end{tabular}
\label{table}
\end{equation}
As can be easily obtained from the expression (\ref{output function}), at
the end of protocol the qubit $B$ is found in the state $\left| \phi
\right\rangle $: the initial state of the qubit \ has hence been restored
deterministically. It is worth noting that the ancilla qubit $B$ is
necessary to restore the initial qubit state.

To implement the restoring machine we adopted polarization encoded qubits by
exploiting the isomorphism between the qubit state $\alpha \left|
0\right\rangle +\beta \left| 1\right\rangle $ and the polarization state $%
\left| \phi \right\rangle _{in}=\alpha \left| H\right\rangle +\beta \left|
V\right\rangle $ of a single photon. The cloning process has been realized
adopting the Quantum Injected Optical Parametric Amplifier $\left(
QIOPA\right) $ as said:\ Figure 2 \cite{DeMa02}. Alice's site is placed on
the $k_{1}$ mode, while Bob's site is placed on the $k_{2}$ mode. Consider
first the case of an input $\overrightarrow{\pi }$-encoded qubit $\left|
\phi \right\rangle _{in}\ $associated with a single photon with wavelength
(wl) $\lambda $, injected on the input mode $k_{1}$ of the $QIOPA$, the
other input mode $k_{2}$ being in the vacuum state. The photon was injected
into a nonlinear (NL) BBO ($\beta $-barium-borate) 1.5 mm thick crystal
slab, cut for Type II phase matching and excited by a sequence of UV\
mode-locked laser pulses having duration $\tau \approx $140 $f\sec $ and wl $%
\lambda _{p}$:\ Figure 2. The relevant modes of the NL 3-wave interaction
driven by the UV pulses associated with mode $k_{p}$ were the two spatial
modes with wave-vector (wv) $k_{i}$, $i=1,2$, each supporting the two
horizontal $(H)$ and vertical $(V)$ {\it linear}-$\overrightarrow{\pi }$'s
of the interacting photons. The $QIOPA$ was $\lambda $-degenerate, i.e. the
interacting photons had the same wl's $\lambda =%
%TCIMACRO{\UNICODE[m]{0xbd}}%
%BeginExpansion
{\frac12}%
%EndExpansion
\lambda _{p}=795nm$. The $QIOPA$ apparatus was arranged in the self-injected
configuration\ shown in Fig. 2 and described in Ref.\cite{DeMa02}. The UV
pump beam, back-reflected by a spherical mirror $M_{p}$ with 100\%
reflectivity and $\mu -$adjustable position ${\bf Z}$, excited the NL
crystal in both directions $-k_{p}$ and $k_{p}$, i.e., correspondingly
oriented towards the right hand side and the left hand side of Fig.2. A
Spontaneous Parametric Down Conversion (SPDC) process excited by the $-k_{p}$
UV\ mode created {\it singlet-states} of photon polarization $(%
\overrightarrow{\pi })$. The photon of each SPDC pair emitted over $-k_{1}$
was back-reflected by a spherical mirror $M$ into the NL crystal and
provided the $N=1$ {\it quantum injection} into the OPA excited by the UV
beam associated with the back-reflected mode $k_{p}$. The twin SPDC\ photon
emitted over mode $-k_{2}$, selected by the devices (Wave-Plate + Polarizing
Beam Splitter: $WP_{T}\ $+ $PBS_{T}$) and detected by $D_{T}$, provided the
''trigger'' of the overall conditional experiment. Because of the EPR
non-locality of the emitted singlet, the $\overrightarrow{\pi }$-selection
made on $-k_{2}$ implied deterministically the selection of the input state $%
\left| \phi \right\rangle _{in}$ on the injection mode $k_{1}$. By adopting $%
\lambda /2$ or $\lambda /4$ wave-plates (WP) with different orientations of
the optical axis, the following $\left| \phi \right\rangle _{in}$ states
were injected:$\;\left| H\right\rangle $, $2^{-1/2}(\left| H\right\rangle
+\left| V\right\rangle )=\left| +\right\rangle $, and $2^{-1/2}(\left|
H\right\rangle +i\left| V\right\rangle )=\left| R\right\rangle $. The three
fixed quartz plates $(Q)$ inserted on the modes $k_{1}$, $k_{2}$ and $-k_{2}$
provided the compensation for the unwanted walk-off effects due to the
birefringence of the NL crystal. An additional walk-off compensation into
the BBO\ crystal was provided by the $\lambda /4$ WP exchanging on mode $%
-k_{1}\ $the $\left| H\right\rangle $ and $\left| V\right\rangle \ 
\overrightarrow{\pi }-$ components of the injected photon. All adopted
photodetectors $(D)$ were equal SPCM-AQR14 Si-avalanche single photon units.
One interference filter with bandwidth $\Delta \lambda =6nm$ was placed in
front of each $D$.

The reversion machine has been realized adopting linear optics and
single-photon detectors (Fig.2). A complete Bell measurement can be realized
adopting a deterministic C-NOT gate\ \cite{Obri03}, while the four Bell
state identification cannot be obtained by simple linear optics elements\ 
\cite{Cals01}. In the present experiment we have restricted our analysis to
the detection of the state $\left| \Psi ^{+}\right\rangle _{SA}$ (\ref
{output function}) realized by means of the polarizing beam splitter $%
PBS_{A} $ and the detector $D_{A}$ and $D_{A}^{\ast }$ (the state $\left|
\Psi ^{-}\right\rangle _{SA}$ is absent since the qubits $S$ and $A$ belongs
to the symmetric subspace). Bob performed the $\sigma _{Z}$ operation by
means of a $\lambda /2$ wave plate.

As first experimental step we have identified the position ${\bf Z}=0$
corresponding to the overlap between the injected photon and the UV pump
pulse (Fig.3-{\bf a}). Let us consider the situation in which we inject the
state $\left| \phi \right\rangle $ and we detect the Bell state $\left| \Psi
^{+}\right\rangle _{SA}.$ When there is a perfect matching between the UV
pump and the injected photon, the polarization state of the photon over mode 
$k_{2}$ should be $\rho _{B}=\left| \phi \right\rangle \left\langle \phi
\right| $, while for ${\bf Z}>>0$ the cloning and restoring machines are
turned off and $\rho _{B}=\frac{1}{2}\left| \phi \right\rangle \left\langle
\phi \right| +\frac{1}{2}\left| \phi ^{\perp }\right\rangle \left\langle
\phi ^{\perp }\right| .$ We have injected the state $\left| R\right\rangle
_{S}$\ and analyzed the output state over the mode $k_{2}$ with a $\lambda
/4 $ waveplate and a polarizing beam splitter $PBS_{2}.$ The output state of
the qubit $B$ was analyzed adopting the couple of detectors $\left\{
D_{2},D_{2}^{\ast }\right\} $; the states $\left| R\right\rangle _{B}$ and $%
\left| L\right\rangle _{B}=2^{-1/2}\left( \left| H\right\rangle _{B}-i\left|
V\right\rangle _{B}\right) $ were, respectively, detected by $D_{2}$ and $%
D_{2}^{\ast }$. The detection of a photon from the detector $D_{T}$ ensured
the injection of the state $\left| \phi \right\rangle _{S}$ over the mode $%
k_{1}$. A coincidence between $D_{A}$ and $D_{A}^{\ast }$ identified the
state $\left| \Psi ^{+}\right\rangle _{SA}$. The coincidence counts $\left(
D_{T},D_{A},D_{A}^{\ast },D_{2}\right) $ and $\left( D_{T},D_{A},D_{A}^{\ast
},D_{2}^{\ast }\right) $ are reported in Fig.3-{\bf a }versus the position $%
{\bf Z}$ of the UV mirror $M_{P}.$ The peak in the coincidence counts of $%
\left( D_{T},D_{A},D_{A}^{\ast },D_{2}\right) $ and the dip in the
coincidence counts of $\left( D_{T},D_{A},D_{A}^{\ast },D_{2}^{\ast }\right) 
$ in correspondence of the matching between the injected qubit and the UV
pump beam are a signature of the realization of the state $\left| \phi
\right\rangle _{B}$.

To completely characterize the output state of the $CM$ + $RM$ process we
have positioned the mirror $M_{P}$ in the position ${\bf Z}=0$ and we have
carried out a single qubit quantum state tomography \cite{Jame01} on the $%
k_{2}$ mode for three different states of the input qubit $\left| \phi
\right\rangle =\left| H\right\rangle $, $\left| \phi \right\rangle =\left|
+\right\rangle $ and $\left| \phi \right\rangle =\left| R\right\rangle $
(Fig.3-{\bf b}) with $\left| \pm \right\rangle =2^{-1/2}\left( \left|
H\right\rangle \pm \left| V\right\rangle \right) $. This analysis is
performed through a $\lambda /4$, a $\lambda /2$, $PBS_{2}$ and the detector 
$D_{2}$ and $D_{2}^{\ast }$. The coincidence counts $\left(
D_{T},D_{A},D_{A}^{\ast },D_{2}\right) $ and $\left( D_{T},D_{A},D_{A}^{\ast
},D_{2}^{\ast }\right) $ are acquired for different settings of the
waveplate positions in order to measure the different Stokes parameters of
the $k_{2}$ mode. The density matrices $\rho _{out}$ in Fig.3-{\bf b} are
represented in the basis $\left\{ \left| \phi \right\rangle ,\left| \phi
^{\perp }\right\rangle \right\} .$ In the ideal case $\rho _{B}=\left| \phi
\right\rangle \left\langle \phi \right| $, as said. To ensure a higher
visibility of the overall process in the measurements, the different modes
selection was assured by narrower pinhole. The experimental results confirms
our theory; the fidelity ${\cal F}_{\phi }=\left\langle \phi \right| \rho
_{out}\left| \phi \right\rangle $ of the overall process are found to be $%
{\cal F}_{H}=0.98\pm 0.01$, ${\cal F}_{+}=0.78\pm 0.01$ and ${\cal F}%
_{R}=0.76\pm 0.01$. The average experimental fidelity of the overall process
has been found to be ${\cal F}=0.84$. This value is found to be largely
above the classical estimation bound achievable by measuring the input
qubit,\ equal to $0.67$.

In conclusions, we have demonstrated experimentally the reversibility of the
cloning-flipping processes by showing how the output of \ these processes
can be exploited to non-locally restore the input qubit $\left| \phi
\right\rangle $ with unit fidelity. `This result, indeed a fundamental one
in the domain of Quantum\ Information, is expected to represent a
significant contribution in modern multipartite quantum cryptography
protocols \cite{Hill99,Fili04,Chen05}. In particular, the adoption of W
entangled states (\ref{Wstate}) for quantum secure communication \cite
{Eibl04,Joo03} and secret sharing protocols belong to the most advanced
issues raised recently in this field. Finally the present experimental
realization is a significant step towards the adoption of the cloning
process for optimal partial state estimation, as recently proposed by \cite
{Cerf05}.

This work has been supported by the FET European Network on Quantum
Information and Communication (Contract IST-2000-29681: ATESIT), by Istituto
Nazionale per la Fisica della Materia (PRA\ ''CLON'')\ and by Ministero
dell'Istruzione, dell'Universit\`{a} e della Ricerca (COFIN 2002).

\centerline{\bf Figure Captions}

\vskip 8mm

\parindent=0pt

\parskip=3mm

Figure.1. General scheme for the cloning machine and the restoring machine.
The restoration of the initial input qubit $\left| \phi \right\rangle $ is
obtained by means of a Bell measurement on the clone qubits $S$ and $A$ at
Alice's site, classical communication, and feedforward operation $\sigma
_{i} $ on the qubit $B$ at Bob's site.

Figure.2. Schematic diagram of the {\it self-injected} Optimal Parametric
Amplifier. The Bell measurement is performed on the cloning mode $k_{1}$
(Alice's site). The photon on mode $k_{2}$ undergoes a $\sigma _{Z}$
operation (Bob's site). The output qubit $S$\ on mode $k_{2}$ is analyzed\
adopting single qubit tomography.

Figure.3. ({\bf a}) Coincidence counts $\left( D_{T},D_{A},D_{A}^{\ast
},D_{2}\right) $ and $\left( D_{T},D_{A},D_{A}^{\ast },D_{2}^{\ast }\right) $
versus the position ${\bf Z}$ of the UV mirror $M_{P}$. Each experimental
point has been measured in a time of $2400s$. ({\bf b})\ Quantum state
tomography of the output qubits $B$ carried out by means of a $\lambda /4$,
a $\lambda /2$, $PBS_{2}$, $D_{2}$ and $D_{2}^{\ast }$ in the position ${\bf %
Z=0}$. For each injected state $\left| \phi \right\rangle $ the
experimentally reconstructed density matrix is represented in the $\left\{
\left| \phi \right\rangle ,\left| \phi ^{\perp }\right\rangle \right\} $
basis. The time required for each matrix reconstruction was $\sim 24$ $h$.

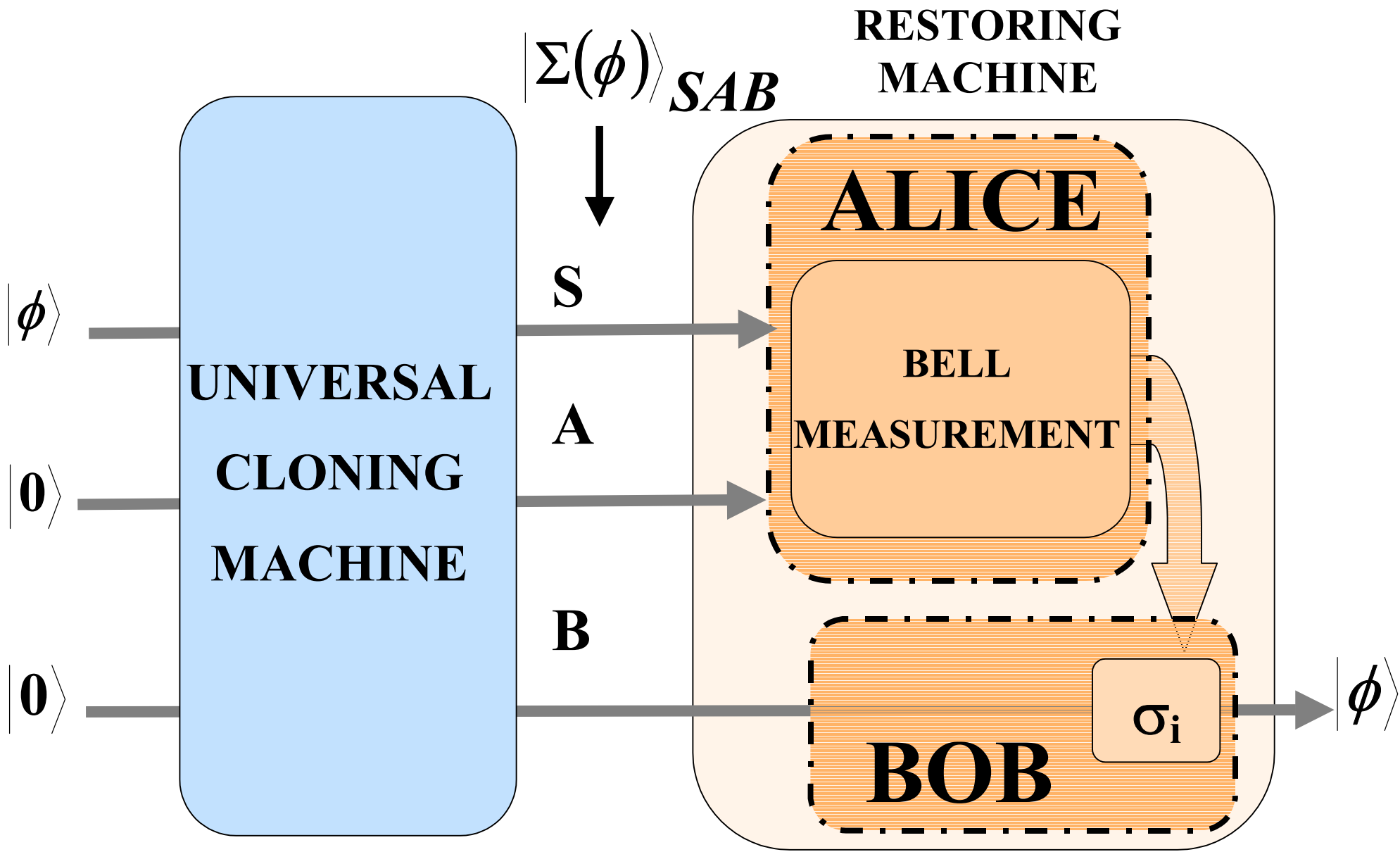
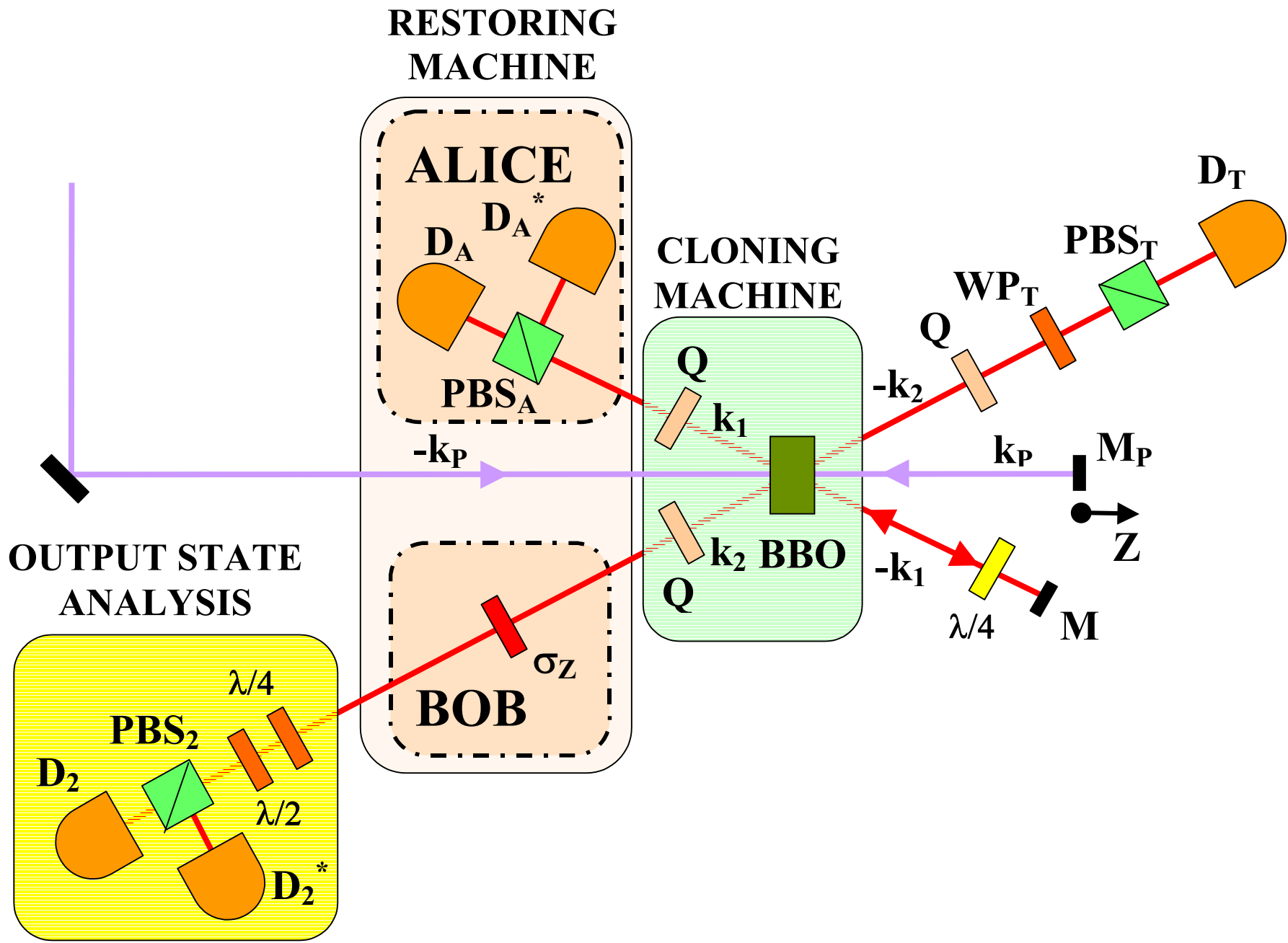
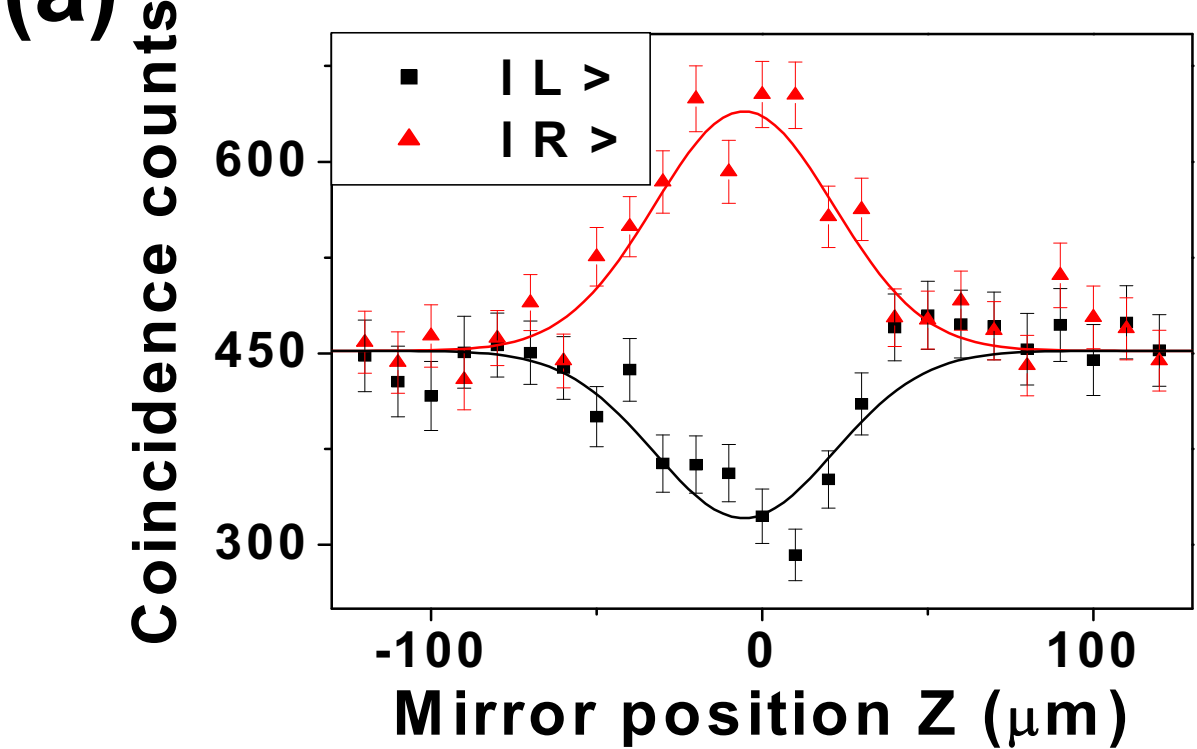
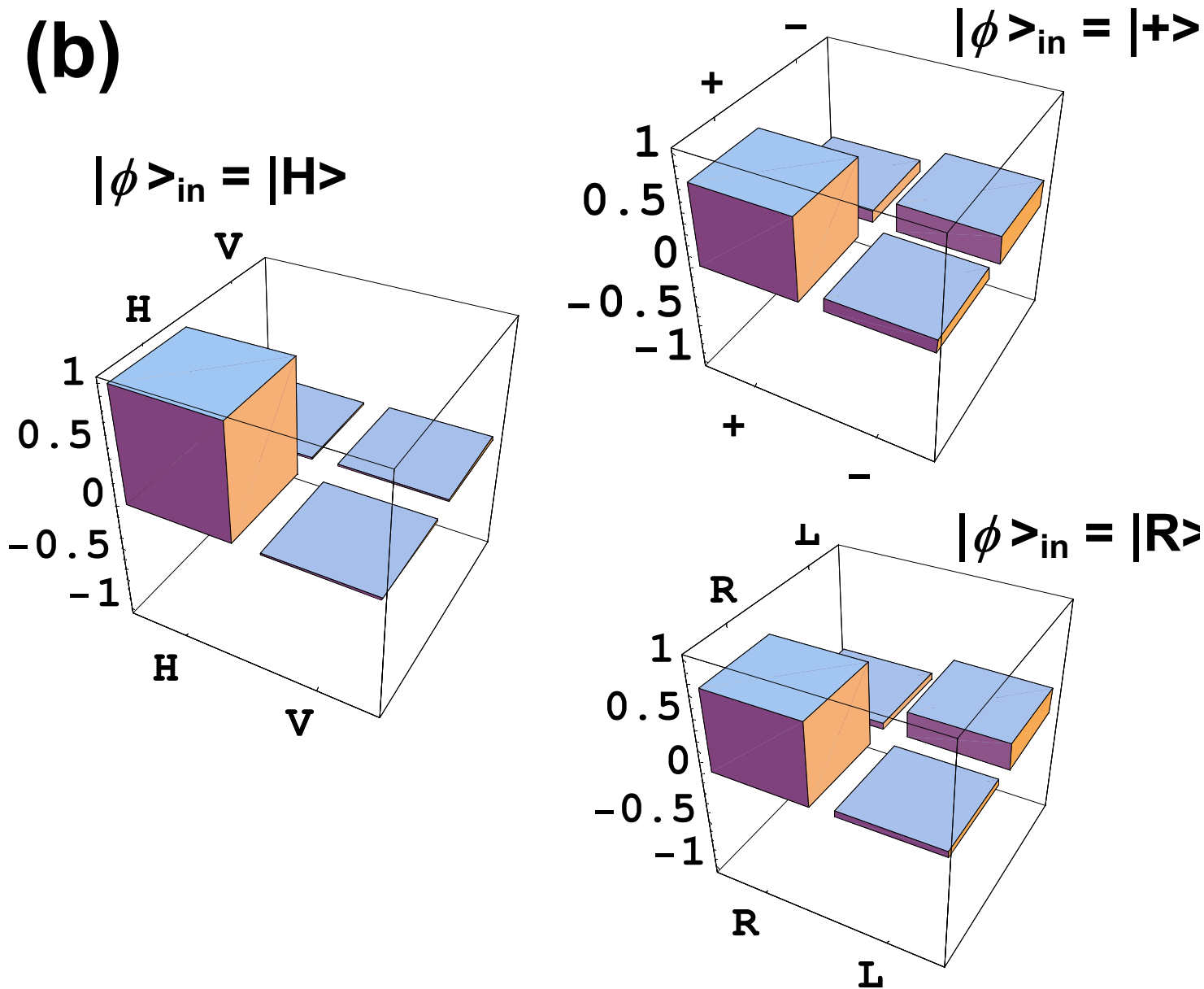
\end{document}